%% file: paper_FC.tex
\documentclass{emulateapj}
\usepackage{savesym}
\savesymbol{tablenum}
\usepackage{siunitx}
\restoresymbol{SIX}{tablenum}
\let\oldendlongtable\endlongtable
\let\oldlongtable\longtable
\makeatletter
\renewenvironment{longtable}[1]{
\oldlongtable{[#1]}
\def\pt@format{#1}
}{\oldendlongtable}
\makeatother
\usepackage{color}
\definecolor{white}{rgb}{1,1,1}

\usepackage{colortbl}
\definecolor{lmugreen}{rgb}{0.01,0.58,0.25}

\usepackage{amsmath}
\usepackage{natbib}
\usepackage{float}
\usepackage{dcolumn}
\usepackage{booktabs}
\usepackage[
	colorlinks=true,
	urlcolor=lmugreen,
	linkcolor=lmugreen,
	citecolor=lmugreen
	]{hyperref}
\usepackage{rotating}
\usepackage{graphicx}
\usepackage{tablefootnote}
\usepackage{longtable}
\usepackage{multirow}
\usepackage[nolist]{acronym}

\usepackage{listings}
\newcommand{\citeorder}[1]{}

\newcommand{\HyperionSec}{\uppercase{H\small yperion}}

\newcommand{\TORUS}{\uppercase{T\small orus}}
\newcommand{\LIME}{\uppercase{L\small ime}}
\newcommand{\MOCASSIN}{\uppercase{M\small ocassin}}

\newcommand{\FCSecF}{\uppercase{F\small lux}}
\newcommand{\FCSecC}{\uppercase{C\small ompensator}}

\DeclareSIUnit\yr{yr}
\DeclareSIUnit\ergs{ergs}
\DeclareSIUnit\Jy{Jy}
\DeclareSIUnit\Hz{Hz}
\DeclareSIUnit\sr{sr}
\DeclareSIUnit\pc{pc}
\DeclareSIUnit\ppm{ppm}
\DeclareSIUnit\kpc{\kilo\pc}
\DeclareSIUnit\AU{AU}
\DeclareSIUnit\eqsign{\text{\ensuremath{=}}}
\DeclareSIUnit\simeqsign{\text{\ensuremath{\simeq}}}
\DeclareSIUnit\mag{mag}
\DeclareSIUnit\gramms{g}
\DeclareSIUnit\gccm{\gramms\per\cubic\centi\metre}
\DeclareSIUnit\microns{\micro\metre}
\DeclareSIUnit\Myr{\mega\yr}
\DeclareSIUnit\Msun{\text{\ensuremath{M_\odot}}}
\DeclareSIUnit\Lsun{\text{\ensuremath{L_\odot}}}
\DeclareSIUnit\Rsun{\text{\ensuremath{T_\odot}}}
\DeclareSIUnit\Minfall{\Msun\per\yr}

\slugcomment{The Astrophysical Journal}

\shorttitle{The FluxCompensator}
\shortauthors{Christine M. Koepferl and Thomas P. Robitaille}

\begin{document}
\title{The FluxCompensator: Making Radiative Transfer Models of hydrodynamical Simulations Directly Comparable to Real Observations}
\author{Christine M. Koepferl$^{1,2}$ and Thomas P. Robitaille$^{1,3}$}

\affil{$^1$ Max Planck Institute for Astronomy, K\"onigstuhl 17, D-69117 Heidelberg, Germany\\
$^2$ University Observatory Munich, Scheinerstr. 1, D-81679 Munich, Germany\\
$^3$ Freelance Consultant, Headingley Enterprise and Arts Centre, Bennett Road Headingley, Leeds LS6 3HN}
\email{koepferl@usm.lmu.de}

\received{22nd June 2017}
\accepted{14th August 2017}

\begin{abstract}
When modeling astronomical objects throughout the universe, it is important to correctly treat the limitations of the data, for instance finite resolution and sensitivity. In order to simulate these effects, and to make radiative transfer models directly comparable to real observations, we have developed an open-source Python package called the \textsc{FluxCompensator} that enables the post-processing of the output of 3-d Monte-Carlo radiative transfer codes, such as \textsc{Hyperion}. With the \textsc{FluxCompensator}, realistic synthetic observations can be generated by modeling the effects of convolution with arbitrary \acp{PSF}, transmission curves, finite pixel resolution, noise and reddening. Pipelines can be applied to compute synthetic observations that simulate observatories, such as the \textit{Spitzer Space Telescope} or the \textit{Herschel Space Observatory}. Additionally, this tool can read in existing observations (e.\,g.~\acs{FITS} format) and use the same settings for the synthetic observations. In this paper, we describe the package as well as present examples of such synthetic observations.
\end{abstract}

\section{Introduction}
\label{C2:intro}
Theoretical simulations (e.\,g.~hydrodynamical simulations) are currently used to reproduce the physical processes in astronomical objects, such as \acp{YSO}, star-forming regions and galaxies. However, we cannot use these simulations directly to predict or explain observational features of such objects. Instead, full radiative transfer calculations need to be performed to properly take into account effects, such as local temperature variations or changes in the optical depth. These radiative transfer calculations provide idealized observations, which are somewhat closer to real observations than the hydrodynamical simulations. In the last decade, the use of idealized observations, from radiative transfer techniques, has been made possible due to the development of full 3-d Monte-Carlo radiative transfer codes (see the extensive review of \citealt{Steinacker:2013}). We list some publicly available examples below:

\begin{itemize}
    \item \textbf{RADMC3D}\\ 
    \textsc{RadMC3d} is a 3-d Monte-Carlo radiative transfer code \citep[for details\footnote{\tiny \url{http://www.ita.uni-heidelberg.de/~dullemond/software/radmc-3d/}}, see][]{DullemondDominik:2004}. It can treat the dust radiative transfer but also gas line radiative transfer for \ac{LTE} and non-\ac{LTE} problems. It has a variety of different geometries implemented and is parallelized. It uses the modified random walk and raytracing.
    \item \textbf{\HyperionSec}\\
    The 3-d dust continuum Monte-Carlo radiative transfer code \textsc{Hyperion} \citep[for details\footnote{\tiny \url{http://www.hyperion-rt.org}}, see][]{Robitaille:2011}, which is fully parallelized and is not dependent on specific synthetic astronomical objects but rather provides a generic set of geometries. \textsc{Hyperion} as other dust radiative transfer codes, treats all dust interactions, such as absorption, re-emission and scattering. \textsc{Hyperion} is an \ac{LTE} code, and, in addition, can treat (simple) non-\ac{LTE} approximations, such as the emission of \acp{PAH}. It uses the Lucy method, the modified random walk, peeling-off and raytracing.
    \item \textbf{\MOCASSIN}\\ 
    \textsc{Mocassin}, a 3-d photoionization and dust non-LTE Monte-Carlo radiative transfer code \citep[for details\footnote{\tiny \url{http://mocassin.nebulousresearch.org}}, see][]{ErcolanoMoccasin:2003,ErcolanoMoccasin:2005,ErcolanoMoccasin:2008}. The code is parallelized and contains a number of grid geometries.
    \item \textbf{\LIME}\\
    \textsc{\LIME} is a molecular excitation and non-LTE spectral line radiation transfer code \citep[for details\footnote{\tiny \url{http://www.nbi.dk/~brinch/lime.php}}, see][]{Brinch:2010} and contains different geometry set-ups and is fully parallelized.
    \item \textbf{\TORUS}\\ p
    \textsc{Torus} is a molecular line and photoionization code which makes use of the 3-d Monte-Carlo radiative transfer technique but also has hydrodynamics, radiation hydrodynamics and self-gravity built-in \citep[for details\footnote{\tiny \url{http://www.astro.ex.ac.uk/people/th2/torus_html/homepage.html}}, see][]{Harries:2000}. It is fully parallelized, has different sets of geometry grids and uses the Lucy method.
\end{itemize}

\begin{figure*}[t]
\includegraphics[width=\textwidth]{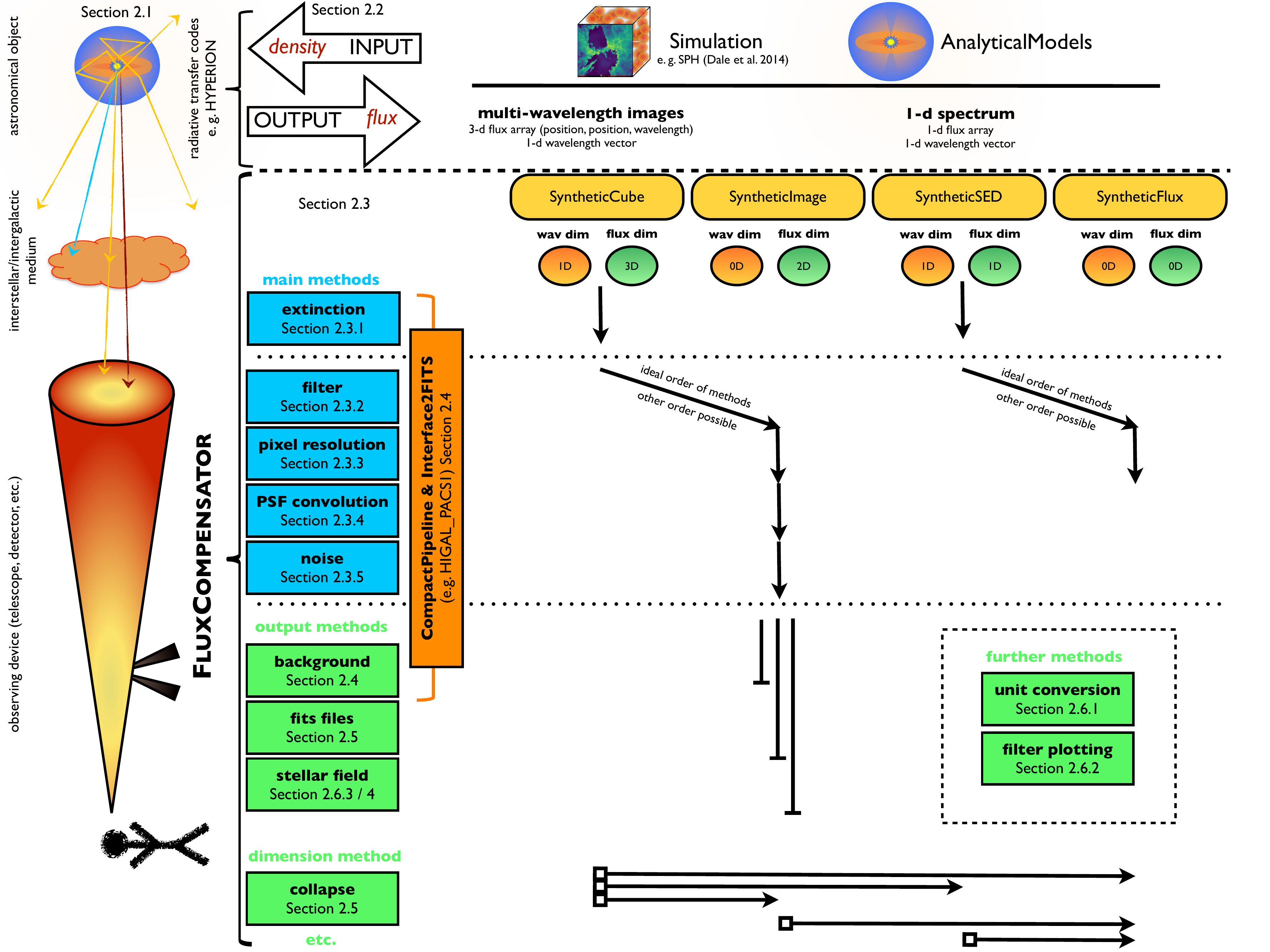}
\caption{Flow scheme of the relevant \textsc{Hyperion} and \textsc{FluxCompensator} input/output structure and the analogy in reality. The relevant sections are highlighted within the scheme.}
\label{C2:Fig:flowchart}
\end{figure*}

\begin{figure*}[t]
\includegraphics[width=\textwidth]{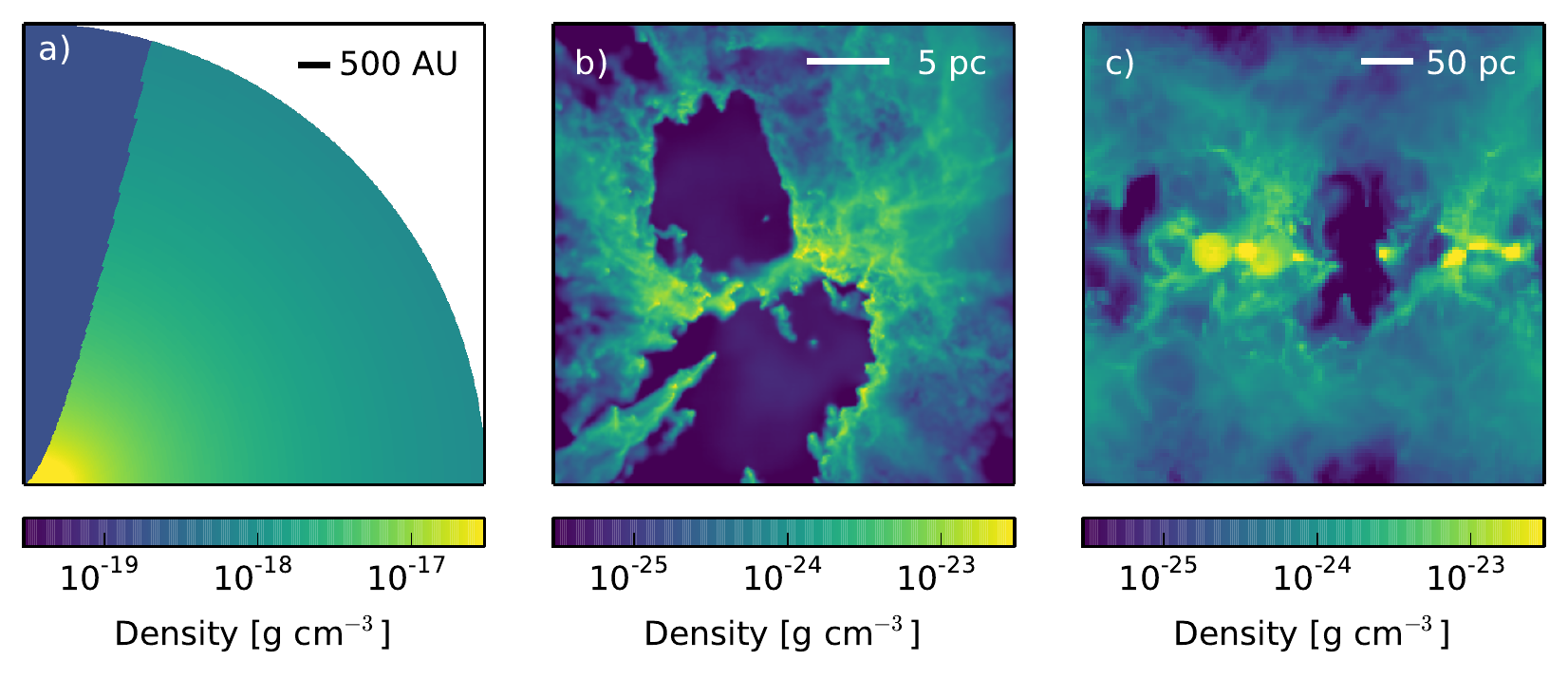}
\caption{Total projected density of a) a \ac{YSO} model, b) a star-forming region, and c) a center of a galaxy.}
\label{C2:Fig:density}
\end{figure*}

Monte-Carlo radiative transfer codes (e.\,g.~\textsc{Hyperion}) can read density structures from the output of theoretical simulations. At every grid point, they sample the probability distributions of a photon being scattered, absorbed or re-emitted, which is propagating through the grid. As a result, the temperature is calculated and flux images can be generated in any given direction or wavelength. For more details, see \citealt{Robitaille:2011} and the extensive review of \citealt{Steinacker:2013}. The radiative transfer, however, only treats the ideal path of the photons from the astronomical object to the telescope. Concretely, with the mere solution of the radiative transfer problem the observational limitations of a telescope, such as finite resolution or the effects on the ideal synthetic observation by optics or detectors within the telescope, have so far not been accounted for. We, hereafter, call such synthetic observations (which lack the treatment of the observational effects) from radiative transfer calculation ``ideal'' synthetic observations.

We have developed a tool, called the \textsc{FluxCompensator}, which treats these limitations and produces "realistic" synthetic observations. Our tool links theoretical studies and radiative transfer modeling with actual astronomical observations by making them directly comparable. The tool was successfully applied in \cite{Koepferl:2015}, \cite{Ercolano:2015} and \cite{Roccatagliata:2015}, as well as in the paper series \textit{"Insights from Synthetic Star-forming Regions"} \cite{KDR1:inprep,KDR2a:inprep,KDR2b:inprep} and the Milky Way Project\footnote{Citizen science project at \url{zooniverse.org} \citep{MWKendrew2016,MWKerton2015,MWBeaumont2014,MWKendrew2012,MWSimpson2012}.}. In this paper, we now present the \textsc{FluxCompensator} tool, which produces these realistic synthetic observations from radiative transfer calculations. In Section~\ref{C2:code}, we describe the code and discuss its compatibility to other radiative transfer codes in Section~\ref{C2:comp}. As an example in Section~\ref{C2:case}, we show realistic synthetic observations of a \ac{YSO} model, a star-forming region and a center of a galaxy and discuss applications and future prospects for the code (Section~\ref{C2:futur}), before summarizing the paper in Section~\ref{C2:summary}. 

\section{Code Overview}
\label{C2:code}
With the \textsc{FluxCompensator} (now available via \url{https://github.com/koepferl/FluxCompensator}, for scheme see Fig.~\ref{C2:Fig:flowchart}) we can post-process radiative transfer calculations, such as from \textsc{Hyperion} \citep{Robitaille:2011}\footnote{Note as the \textsc{FluxCompensator} was designed as an post-processing tool for \textsc{Hyperion} however it can be used by other codes. Therefore, the reader should see the \textsc{Hyperion} \citep{Robitaille:2011} reference as an example of many other radiative transfer codes. For the remainder of this paper, however, we will use \textsc{Hyperion} as a place holder for other radiative transfer codes.}, by reading in the radiative transfer output and using these, to produce realistic synthetic observations. The \textsc{FluxCompensator} follows the same philosophy as \textsc{Hyperion} in using object-oriented Python scripts and making it therefore more user friendly and generic. Very little Python knowledge is required to work with either \textsc{Hyperion} or the \textsc{FluxCompensator}. For the compatibility with other radiative transfer codes see Section~\ref{C2:comp}. 

For the comprehension of the following sections Figure~\ref{C2:Fig:flowchart} is helpful where we visualized the methodology of the \textsc{FluxCompensator} and the concept of synthetic observation in a flow scheme. The relevant sections a highlighted within the scheme.

\subsection{Life of a Real Photon}
\label{C2:Sec:life_real_photon}
In nature, photons emitted from a source are scattered, absorbed and re-emitted by matter (dust and gas) in astronomical objects (e.\,g.~\acp{YSO}, star-forming regions, galaxies). These photon processes happen many times as they propagate through matter. Absorption of photons by dust and gas causes the matter to heat up. The emitted light from the astronomical object (in form of photons) is reddened by interstellar extinction. Further only some of the emitted photons reach the telescope, because of its finite opening angle. Photons of different energy enter the telescope. The beam of light diffracts at the opening of the telescope. The diffraction pattern is described by the \ac{PSF} of the telescope\footnote{In a non-ideal observational system also atmospheric matter (e.\,g.~seeing) which causes the \ac{PSF} to blur.  For space observatories no seeing exists while the shapes of the \acp{PSF} depend merely on the characteristics of the instruments.  The angular resolution of the telescope with diameter $D$ relates to $\lambda/D$. Note that the above is strictly only true infrared light or shorter wavelengths not for the radio emission, where the distortion is called beam and can no longer be approximated by an Airy disc. }. The spectral transmission\footnote{Note that when we speak of spectral transmission of the "telescope and detector" we refer to all effects introduced by parts of the observing device (e.\,g.~filters, telescope configuration, dispersive elements) that select wavelength specific photons. This filtering introduces flux loss, also  in the non-selected wavelength areas of the spectrum. Only when the "spectral transmission function" is not a normalized box function the flux loss in the selected wavelength range is zero.} of the telescope further filters the radiation in such a way that only photons within a certain wavelength range can reach the detector. 

The light of the photons detected can either be measured by a single value -- the photometric flux which is a scalar -- or by spatially resolving the light -- photometric images, which are essentially 2-d flux arrays. If we measure photometric fluxes for different filters, instruments, or telescopes, these can then be combined into 1-d spectrum\footnotemark[10]. Observational images are produced by exposing many pixels in a two-dimensional configuration to the photons. The pixels map the projected positions, of the last location where the photons interacted in the observed object, onto a 2-d plane\footnotemark[11]. The images are also altered by the \ac{PSF} of the telescope and furthermore, the pixel resolution of the detector has an important effect on the images.
\footnotetext{1-d flux and wavelength vector}
\footnotetext{2-d flux array (position, position) and scalar wavelength}

\subsection{Life of a Synthetic Photon in \HyperionSec}
\label{C2:Sec:life_syn_photon}

\textsc{Hyperion} is a 3-d dust continuum Monte-Carlo radiative transfer tool that can take any arbitrary 3-d distribution of dust and sources, and compute the temperature structure as well as idealized multi-wavelength observations. 

It can be used to compute \textsc{AnalyticalModels} (toy models with an analytical density distribution), such as \acp{YSO} (see Figure~\ref{C2:Fig:density}\textcolor{lmugreen}{(a)}), but it can also read in any arbitrary 3-d density distributions from hydrodynamical simulations (see Figure~\ref{C2:Fig:density}\textcolor{lmugreen}{(b,c)}). Figure~\ref{C2:Fig:density}\textcolor{lmugreen}{(b)} shows a time-step in an \ac{SPH} simulation of a \SI{e4}{\Msun} star-forming region by \cite{DaleIoni:2012}. In this particular run only ionizing sources contribute to the stellar feedback. In Figure~\ref{C2:Fig:density}\textcolor{lmugreen}{(c)}, we see the surface density of a center of a galaxy extracted from the simulations performed by the \textsc{SILCC} collaboration \citep{Girichidis:2015,Walch:2014}.

\textsc{Hyperion} samples \acp{PDF} for the photon interactions and computes the temperatures in the grid. The photons interact with the dust and eventually can escape the system. The photons escaping in the direction of the telescope are then used to produce the synthetic observations. Depending on the initial set-up, 1-d spectrum\footnotemark[10] and/or multi-wavelength images\footnotemark[12] are produced. \textsc{Hyperion} scales these ideal synthetic observations to a certain distance and converts these to the requested units once they are extracted. For more details see \cite{Robitaille:2011}.
\footnotetext{3-d flux array (position, position, wavelength) and 1-d wavelength vector}

\subsection{Virtual Pipeline of the \FCSecF\FCSecC}
\label{C2:virt}
The \textsc{FluxCompensator} uses 3-d spectral cubes\footnotemark[12] or 1-d spectrum\footnotemark[10] from radiative transfer calculations and simulates the effects which are introduced by the telescope. In order to make the \textsc{FluxCompensator} user-friendly, the framework itself can support any transmission curve and \ac{PSF} file provided by the user. Furthermore, we have included common extinction laws, \ac{PSF} files, spectral transmission curves and further information in the database of the \textsc{FluxCompensator}. For references and the extent of the information in the database, see Table~\ref{C2:Appendix:Tab} in Appendix~\ref{C2:Appendix}. It is left up to the users whether they use this database or whether they just use the methods of the \textsc{FluxCompensator} but load their preferred resources with a separate line of code instead.

The dimensions of the observations in flux change when passing through the telescope and the detectors. For example, multi-wavelength photons that enter the telescope can be represented by an idealized 3-d spectral cube. The spectral transmission of the telescope and detector, filters only photons within certain slices of this spectral cube and weights them, so that a two-dimensional image remains. The \textsc{FluxCompensator} therefore has four different classes to treat the different dimensions:
\texttt{SyntheticCube}, \texttt{SyntheticImage}, \texttt{SyntheticSED}, \texttt{SyntheticFlux}. Each has a different flux and wavelength dimension, as highlighted in Table~\ref{C2:Tab:classes}.
\begin{table}[t]
	\caption{Class Dimensions in the \textsc{FluxCompensator}}
	\label{C2:Tab:classes}
		\begin{center}
		\begin{minipage}{\textwidth}
			\begin{tabular*}{0.48\textwidth}{p{0.17\textwidth}p{0.17\textwidth}p{0.1\textwidth}}
                \hline\\[-13pt]
                \hline\\[-5pt]
                &&\\
				\multicolumn{1}{l}{class} 	&	\multicolumn{1}{l}{flux}	&	\multicolumn{1}{l}{wavelength}	\\[6pt]
 				\hline\\[-5pt]
\texttt{SyntheticCube}	&	3-d (x, y, wav)	&	1-d array			\\
\texttt{SyntheticImage}	&	2-d (x, y)     	&	0-d array			\\
\texttt{SyntheticSED}	 	&	1-d (wav)      	&	1-d array			\\
\texttt{SyntheticFlux}	&	0-d (wav)	     	&	0-d array			\\[6pt]
 \hline\\[-5pt]
  			\end{tabular*}
  			\end{minipage}
		\end{center}
\end{table}
The methods of the \textsc{FluxCompensator} can be called in any arbitrary order as long as the dimensions allow it (i.\,e.~one cannot apply a filter transmission curve to a 2-d image). The order might affect the speed of the numerical calculation, while producing a comparable physical result. In the following sections, we will present one scenario of using the methods of the \textsc{FluxCompensator} which simulates the effects, such as those introduced by a telescope and its detector. Pipelines with a different order of the methods (e.\,g.~extinction, PSF convolution, noise) presented in Sections~\ref{C2:ext} to Sections~\ref{C2:noise} are also supported. Not all methods need to be applied and the user can select those appropriate for the specific scientific question.
\begin{figure*}[t]
\includegraphics[width=\textwidth]{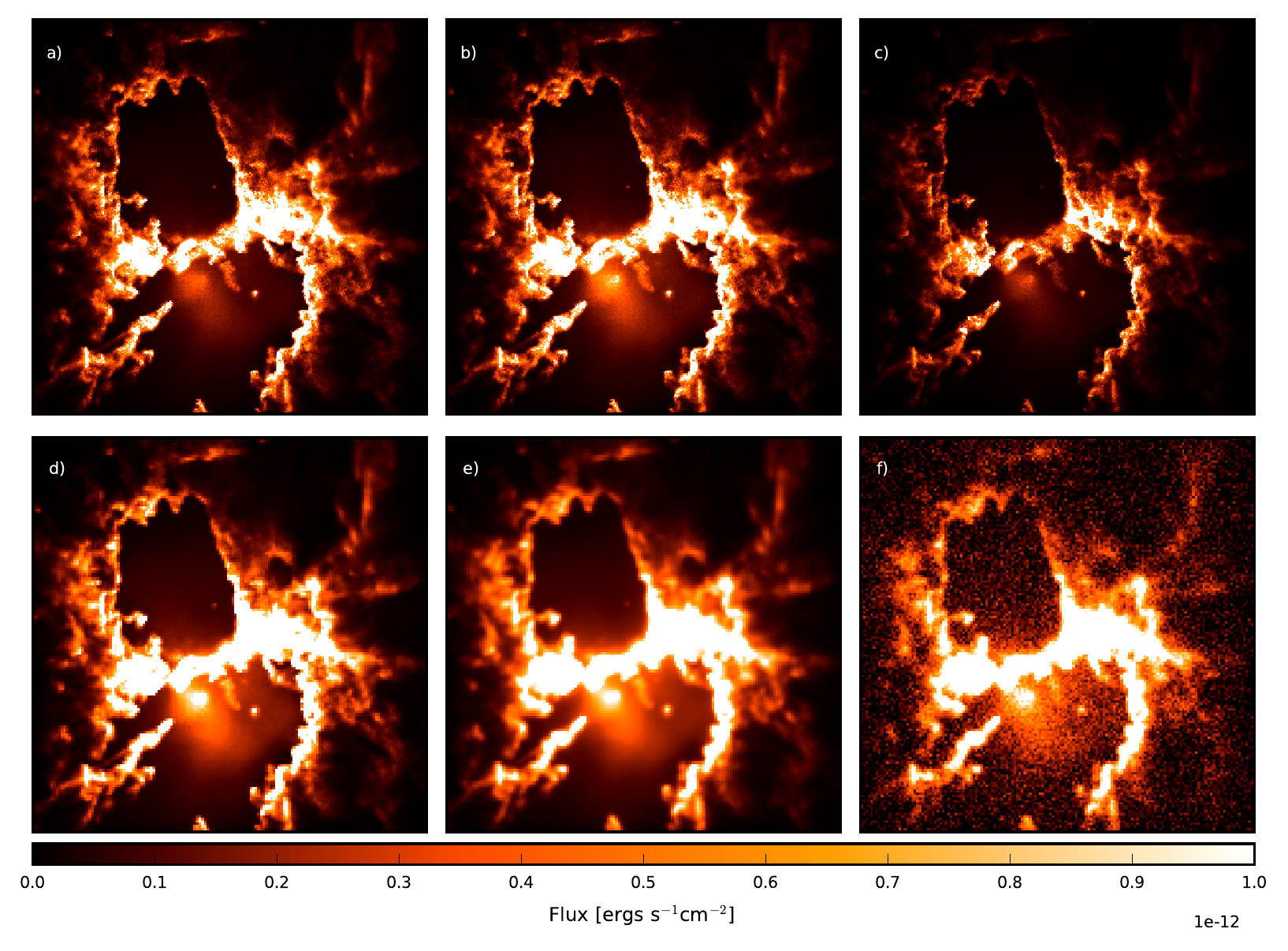}
\caption{Snapshots during the ongoing pipeline: a) \textsc{Hyperion} output image slice closest to $\lambda=\SI{70}{\microns}$, b) image after convolution with \acs{PACS} \SI{70}{\microns} transmission curve, c) \acs{PACS} \SI{70}{\microns} image after reddening, d) \acs{PACS} \SI{70}{\microns} image with corresponding pixel resolution, e) \acs{PACS} \SI{70}{\microns} image after \acs{PACS} \ac{PSF} convolution, f) \acs{PACS} \SI{70}{\microns} with Gaussian noise.}
\label{C2:Fig:pipeline}
\end{figure*}

From Section~\ref{C2:ext} to Section~\ref{C2:noise}, we will illustrate the transformation within the pipeline of the synthetic observations from ideal to realistic, by using snapshots displayed in Figure~\ref{C2:Fig:pipeline}. Figure~\ref{C2:Fig:pipeline}\textcolor{lmugreen}{(a)} shows the ideal synthetic observation slice extracted from the \textsc{Hyperion} output closest to \SI{70}{\microns}. 

\subsubsection{Extinction}
\label{C2:ext}
The photons emitted from the astronomical objects do not usually reach the observing telescope unaffected by extinction (between the object and telescope). By accounting for interstellar reddening \citep[see][]{Carroll:2014,Carroll:1996} the observations are made more realistic. 
The \textsc{FluxCompensator} supports loading arbitrary extinction laws (or provides the law\footnote{Note that this extinction law breaks down above \SI{1}{\milli\meter}.} of \citealt{Kim:1994} in the database).   

With the opacity values $k_\lambda$ provided by the extinction law, we can estimate the wavelength-dependent extinction $A_\lambda$ with
\begin{eqnarray}
A_\lambda = A_V \cdot \frac{k_\lambda}{k_V},
\end{eqnarray}
where the optical extinction $A_V$ is a free input parameter, $k_V$ is interpolated by the \textsc{FluxCompensator} from $k_\lambda$ at optical wavelengths with linear interpolation \citep[see][]{NumericalRecipes}. We redden the flux with 
\begin{eqnarray}
F_{\textup{extincted}}(\lambda) = F_{\textup{intrinsic}}(\lambda) \cdot \num{10}^{\num{-0.4} \cdot A_\lambda}.
\end{eqnarray}
We show the reddened Figure~\ref{C2:Fig:pipeline}\textcolor{lmugreen}{(b)} \ac{PACS} \SI{70}{\microns} synthetic image in Figure~\ref{C2:Fig:pipeline}\textcolor{lmugreen}{(c)}. We used the extinction law from the \textsc{FluxCompensator} database and assumed an optical extinction of $A_V=\num{500}$. At \SI{70}{\microns}, the effect of extinction is negligible. Therefore, realistic values e.\,g.~$A_V=\num{20}$, would make Figure~\ref{C2:Fig:pipeline}\textcolor{lmugreen}{(c)} indistinguishable from Figure~\ref{C2:Fig:pipeline}\textcolor{lmugreen}{(b)}. For now, for illustration purposes we adapt this "unphysical" value.

Note that the reddening of the observation is due to interstellar or intergalactic extinction and is not strictly an instrumental effect as remaining effects which can be simulated with the \textsc{FluxCompensator}.

\subsubsection{Spectral Transmission Convolution}
\label{C2:filter}
When photons of different energies pass through a detector in a real telescope, their different energies (or wavelengths) are weighted by the transmission of the whole system -- optics, filter, camera \citep[see][]{,Carroll:2014,Carroll:1996}. In the \textsc{FluxCompensator} an algorithm performs the rebinning of the spectral transmission and the convolution with the multi-wavelength images or 1-d spectrum. The resulting 2-d \texttt{SyntheticImage} (of an initial 3-d \texttt{SyntheticCube}) or a 0-d \texttt{SyntheticFlux} (of an initial 1-d \texttt{SyntheticSED}) represent the flux observed by the detector. 

The tool can either load arbitrary transmission curves with an extra line of code or use the transmission curves from the database. In Figure~\ref{C2:Fig:pipeline}\textcolor{lmugreen}{(b)}, we present the synthetic image, where Figure~\ref{C2:Fig:pipeline}\textcolor{lmugreen}{(a)} was weighted with a \acs{PACS} \SI{70}{\microns} filter from the \textsc{FluxCompensator} database. The database currently has 24 transmission curves from \ac{2MASS}, \emph{Spitzer}, \emph{Herschel}, \ac{IRAS} and \ac{WISE} with central wavelengths from \SIrange{1.235}{500}{\microns}. Table~\ref{C2:Appendix:Tab} in Appendix~\ref{C2:Appendix} lists the information about the filter transmissions and their references.

For arbitrary loaded filters, two parameters are passed to the \textsc{FluxCompensator} which represent the nature of the filter \citep{Robitaille:2007}. The first parameter is $\alpha$, which resembles the taken assumptions in order to convert the flux into the monochromatic flux density $F_\nu$: 
\begin{eqnarray}
\nu^\alpha F_\nu = const.
\end{eqnarray}
The transmission in a filter curve file $R_{\nu,\textup{input}}$ goes such as 
\begin{eqnarray}
R_{\nu,\textup{input}} = R_\nu\nu^\beta
\end{eqnarray}
to the filter transmission $R_\nu$. The units of the transmission curve are set by the exponent $\beta$. When the transmission curve is given as a function of photons $\beta = \num{0}$, while $\beta = \num{-1}$ when the transmission is given as a function of energy. See Table~\ref{C2:Appendix:Tab} in Appendix~\ref{C2:Appendix} for $\alpha$ and $\beta$ values of the filters in the database and their references.

To weight the synthetic data with the spectral transmission, the \textsc{FluxCompensator} calculates the effective spectral response \citep{Robitaille:2007} at every wavelength slice of the 3-d \texttt{SyntheticCube} or 1-d \texttt{SyntheticSED}:
\begin{eqnarray}
R_{\nu,\textup{response}} = \frac{1}{\nu_0^\alpha}\frac{R_{\nu,\textup{input}}/\nu^{1+\beta}}{\int R_{\nu,\textup{input}}/\nu^{1+\alpha + \beta}},
\end{eqnarray}
where $\nu_0$ is the central frequency of the filter. The sum of the weighted flux with the spectral response then gives the filtered flux:
\begin{eqnarray}
F_{\nu_0,\textup{filtered}} = \sum\limits_{\nu=\nu_{min}}^{\nu_{max}} R_{\nu,\textup{response}} F_\nu,
\end{eqnarray}
within the limits of the filter $\nu_{min}$ and $\nu_{max}$. For more details see the Appendix of \cite{Robitaille:2007}.

\subsubsection{Pixel Resolution}
\label{C2:res}
The pixel resolution of the initial ideal synthetic observation is determined by the set-up of the radiative transfer model and by the distance, which is set by the user in the \textsc{Hyperion} model. The resulting pixel resolution remains constant until actively changed. From Figure~\ref{C2:Fig:pipeline}\textcolor{lmugreen}{(a)} to Figure~\ref{C2:Fig:pipeline}\textcolor{lmugreen}{(c)} the images are still in the intrinsic resolution of about \ang{;;2}. 

However, with the \textsc{FluxCompensator}, the user can also change the pixel resolution after the radiative transfer calculation. This might be useful for example to compare the synthetic observations with a real observation from a certain detector with a specific pixel resolution. This is especially necessary if \ac{PSF} convolution with a telescope-specific \ac{PSF} file is anticipated. 

We have designed an algorithm which redistributes the flux from the initial pixels (within \texttt{SyntheticCube} and \texttt{SyntheticImage}) to a new grid of different size and which causes a change in pixel resolution. Its efficiency is of ${\mathcal O}(N^2)$ and is flux and brightness conserving. 

In Figure~\ref{C2:Fig:pipeline}\textcolor{lmugreen}{(d)}, the synthetic image of Figure~\ref{C2:Fig:pipeline}\textcolor{lmugreen}{(c)} was rescaled to a pixel resolution of \ang{;;4}, which is a specific pixel resolution used for \acs{PACS} \SI{70}{\microns} images.

\subsubsection{PSF Convolution}
\label{C2:psf}
The effects of diffraction \citep[see][]{Carroll:2014,Carroll:1996,Feynman1} at the opening of the observing telescope represent one of the largest effects which distinguishes real from ideal observations\footnote{Note that this strictly only effects  infrared and shorter wavelengths not the radio emission.}. The \textsc{FluxCompensator} supports currently three different approaches of convolving \citep[see][]{Bronstein,NumericalRecipes} an image (within \texttt{SyntheticCube} and \texttt{SyntheticImage}) with a \ac{PSF}: 

\begin{itemize}
\item PSF File\\
It is a good choice to convolve the image with a \ac{PSF} file which was provided in the telescope's documentation. Any arbitrary \ac{PSF} file is supported and can be easily added (with one additional line of code) and a \texttt{FilePSF} object can be constructed. Instances of the \texttt{FilePSF} class for the {\it Spitzer Space Telescope} and {\it Herschel Space Observatory} are already stored in the database of the \textsc{FluxCompensator}. In Figure~\ref{C2:Fig:pipeline}\textcolor{lmugreen}{(e)}, the synthetic image of Figure~\ref{C2:Fig:pipeline}\textcolor{lmugreen}{(d)} was convolved with the \acs{PACS} \SI{70}{\microns} \texttt{FilePSF} class provided in the database. Currently, the database has 28 \texttt{FilePSF} classes constructed by \ac{PSF} files from \emph{Spitzer} (\ac{IRAC}, \ac{MIPS}) and \emph{Herschel} (\ac{PACS}, \ac{SPIRE}). They were obtained from the documentation for the various telescopes (for references see Table~\ref{C2:Appendix:Tab} in Appendix~\ref{C2:Appendix}).

\item Gaussian\\
Diffraction at a circular opening of an optical or infrared telescope with diameter $d$ is mathematically expressed by a \ac{PSF} profile which is an Airy function \citep[see][]{,Carroll:2014,Carroll:1996}. To a  first order approximation this profile can be represented by a 2-d Gaussian. For an image with pixel resolution $\theta$ in radians per pixel, the standard deviation $\sigma_{\textup{Gauss}}$ of the Gaussian profile \citep[see][]{Bronstein} is given by
\begin{eqnarray}
\sigma_{\textup{Gauss}} = \frac{1}{2\sqrt{2\ln2}} \frac{\lambda}{d~\theta}
\end{eqnarray}
in units of pixels. The class \texttt{GaussianPSF} within the \textsc{FluxCompensator} then sets up this \ac{PSF} and a method convolves the image with this Gaussian profile.

\item Arbitrary Function\\
In some cases it might be reasonable to convolve the image with an array constructed from an arbitrary function. The \textsc{FluxCompensator} supports this with the class \texttt{FunctionPSF}. 
\end{itemize}

\begin{figure*}[t]
\includegraphics[width=\textwidth]{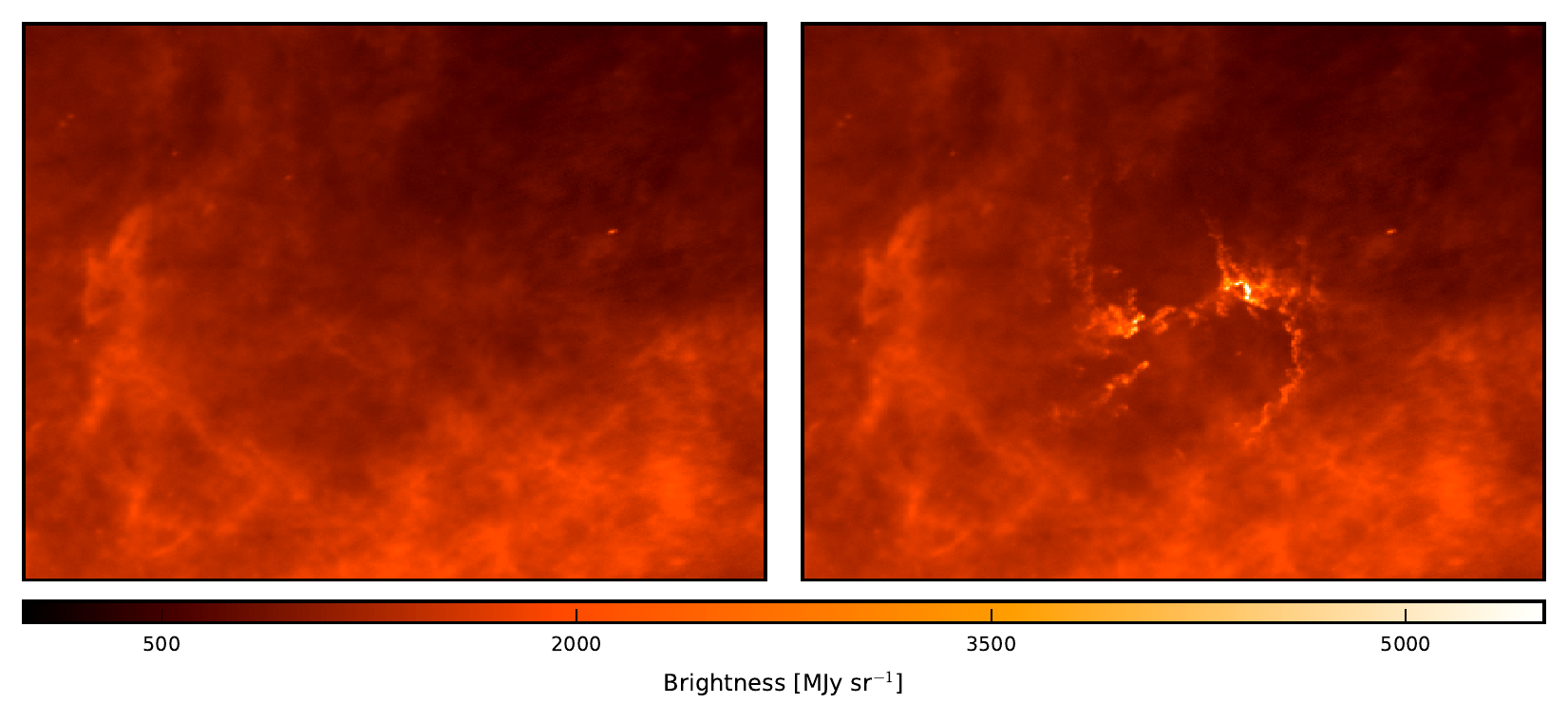}
\caption{Before/after visualization of the realistic synthetic observation (right) of a star-forming region added to a \ac{PACS} \SI{70}{\microns} \ac{Hi-GAL} observation (left/right).}
\label{C2:Fig:add2obs}
\end{figure*}

In any case, it is important to adjust the pixel resolution to the observing detector's pixel resolution before convolving with any \ac{PSF} class described above (see Section~\ref{C2:res}). Further caution is necessary if the class \texttt{FilePSF} is applied on \texttt{SyntheticCube}. A single \ac{PSF} may not be applicable to all wavelengths in the spectral cube. Therefore, we recommend to first convolve with a filter (see Section~\ref{C2:filter}) and then convolve with \texttt{FilePSF}, which is also numerically faster. In future we may implement a wavelength-dependent \ac{PSF}.

\subsubsection{Noise}
\label{C2:noise}
While observing, the statistical noise in an image depends on properties, such as the exposure time and/or the noise contribution from the reading of the camera grid \citep{Howell:2006}.

The \textsc{FluxCompensator} provides a random Gaussian noise method, which adds noise to the images (images in \texttt{SyntheticCube} and \texttt{SyntheticImage}). The input values are the mean $\mu_{\textup{noise}}$ and the standard deviation $\sigma_{\textup{noise}}$ (in the units of the image) of the Gaussian random distribution \citep[see][]{NumericalRecipes}. 

It is possible that in future the \textsc{FluxCompensator} will be able to estimate the standard deviation directly from the information provided about the specific detector in the built-in database. In Figure~\ref{C2:Fig:pipeline}\textcolor{lmugreen}{(f)}, a noise of level $\sigma_{\textup{noise}}=\SI[mode=text]{7e-14}{ergs.s^{-1}.cm^{-2}}$ has been assumed for illustration purposes.

\subsubsection{Example Pipeline}
\label{C2:example}
In the following, we give an example of what a virtual pipeline looks like: We load the Python modules (line 1 to 5), read in the \textsc{Hyperion} output (line 7 to 8), start the \textsc{FluxCompensator} (line 10 to 11), account for filter convolution (line 13 to 18), extinction (line 20 to 21), pixel resolution (line 23 to 24), \ac{PSF} convolution (line 26 to 31) and noise (line 33 to 34). 
\begin{lstlisting}[basicstyle=\ttfamily\footnotesize\fontsize{12}{8}, language=Python]
import numpy as np

from hyperion.model import ModelOutput
from hyperion.util.constants import pc, kpc
from fluxcompensator.cube import *

m = ModelOutput('hyperion_output.rtout')
rt_model = m.get_image(group=0, inclination=0, 
                       distance=10 * kpc, 
                       units='ergs/cm^2/s')

# initial SyntheticObservation array
c = SyntheticCube(input_array=rt_model, 
                  unit_out='ergs/cm^2/s',
                  name='test_cube')

# call object from the filter database
import fluxcompensator.database.missions as filters
filter_input = filters.PACS1_FILTER

# convolve with filter
filtered = c.convolve_filter(filter_input)

# dered with provided extinction law
ext = filtered.extinction(A_v=20.)

# change pixel resolution to 
# PACS1 pixel resolution
zoom = ext.change_resolution(
           new_resolution=
           filters.PACS1_PSF_RESOLUTION)

# call object from the psf database
import fluxcompensator.database.missions as PSFs
psf_object = PSFs.PACS1_PSF

# convolve with PSF
psf = zoom.convolve_psf(psf_object)

# add noise
noise = psf.add_noise(mu_noise=0, 
                      sigma_noise=7e-14)
\end{lstlisting}
In 42 lines of code we have shown how to create a realistic synthetic observation (here in \acs{PACS} \SI{70}{\microns}). 

\subsection{Compact Pipelines and Synthetic Interface to Observation Files}
\label{C2:main}

Since the \textsc{FluxCompensator} (now available via \url{https://github.com/koepferl/FluxCompensator}) is constructed as an Python module, it is possible to construct and loop over entire pipelines, such as the example of Section~\ref{C2:example}, and to generate realistic synthetic images for different detectors and telescopes. The feature \texttt{CompactPipeline} in the \textsc{FluxCompensator} can be used to construct a wrapping code around an existing pipeline. 

\texttt{CompactPipeline} can pass a mixture of self-constructed \ac{PSF} or Filter classes and/or usage of the database and classes such as \texttt{GaussianPSF}. The \textsc{FluxCompensator} supports the construction of these compact pipelines but also provides pre-constructed compact pipelines of the major infrared continuum surveys \ac{2MASS}, \ac{GLIMPSE}, \ac{WISE}, \ac{MIPSGAL}, \ac{Hi-GAL}: 
\begin{itemize}
\item \ac{2MASS}: \texttt{TWOMASSSURVEY\_J}, \texttt{TWOMASSSURVEY\_H}, \texttt{TWOMASSSURVEY\_K}\\[-0.5cm]
\item \ac{GLIMPSE}: \texttt{GLIMPSE\_IRAC1}, \texttt{GLIMPSE\_IRAC2}, \texttt{GLIMPSE\_IRAC3}, \texttt{GLIMPSE\_IRAC4}\\[-0.5cm]
\item \ac{MIPSGAL}: \texttt{MIPSGAL\_MIPS1}, \texttt{GLIMPSE\_MIPS2}\\[-0.5cm]
\item \ac{WISE}: \texttt{WISESURVEY\_WISE1}, \texttt{WISESURVEY\_WISE2}, \texttt{WISESURVEY\_WISE3}, \texttt{WISESURVEY\_WISE4}\\[-0.5cm]
\item \ac{Hi-GAL}: \texttt{HIGAL\_PACS1}, \texttt{HIGAL\_PACS3}, \texttt{HIGAL\_SPIRE1}, \texttt{HIGAL\_SPIRE2}, \texttt{HIGAL\_SPIRE3}\\[-0.5cm]
\end{itemize}

Further, the synthetic interface class, called \texttt{Interface2FITS}, acts as an interface between observations (e.\,g.,~\acl{FITS} -- or \acs{FITS} -- files) and the ideal radiative transfer output. It supports the direct comparison of real observations stored in \acs{FITS} files. The \textsc{FluxCompensator} reads information from the header of the \acs{FITS} file in order to produce realistic synthetic observations with the same wavelength, reddening, pixel resolution and \ac{PSF} specific for telescope and detector of the real observation. Information, such as the distance, the optical extinction coefficient of the object and the exposure time and/or noise contribution, are input parameters. It is also possible to load detector and telescope information which are not available in the built-in database.

When using the built-in interface little knowledge of Python programming is required. We show an example of the \textsc{FluxCompensator} interface code: the Numpy \citep{NumPy}, \textsc{FluxCompensator} and \textsc{Hyperion} modules are loaded (line 1 to 5); the radiative transfer spectral cube is read in (line 7 to 8); the realistic synthetic observation is produced and saved to a \acs{FITS} file (line 10 to 12). It is further possible to combine the realistic synthetic observation with the original \acs{FITS} file of the real observation at a given location (line 13). When comparing the synthetic observation directly with an astronomical object the \texttt{Interface2FITS} might be helpful since it replaces the background estimation. 

\begin{lstlisting}[basicstyle=\ttfamily\footnotesize\fontsize{12}{8}, language=Python]
import numpy as np
from hyperion.model import ModelOutput
from hyperion.util.constants import kpc
from fluxcompensator.interface import Interface2FITS
from fluxcompensator.database.pipeline import HIGAL_PACS1

m = ModelOutput('hyperion_output.rtout')
rt_model = m.get_image(group=0, inclination=0, 
                       distance=10*kpc,
                       units='ergs/cm^2/s')

fluxcompensator = Interface2FITS(
             obs='real_obs_pacs70.FITS', 
             model=rt_model,
             pipeline=HIGAL_PACS1, 
             exposure=10, 
             A_v=20)

fluxcompensator.save2FITS('synthetic_obs')
fluxcompensator.add2observation('add2obs', 
                       position_pix=(3000,2500))
\end{lstlisting}

In Figure~\ref{C2:Fig:add2obs}, we see the output of this code when starting with the radiative transfer model constructed from the density distribution provided by \ac{SPH} simulations of \cite{DaleIoni:2012}. Compare this result with Figure~\ref{C2:Fig:density}\textcolor{lmugreen}{(b)} and note how well this synthetic star-forming region blends into the \ac{Hi-GAL} \citep{Molinari:2010} background.

\subsection{Pipeline Post-processing \& Outputs}
When starting from a 3-d scalar cube from \textsc{Hyperion}, it might be interesting to inspect the current \ac{SED} at some steps in the pipeline. With the \textsc{FluxCompensator}, it is possible at every step in the pipeline to convert a 3-d spectral cube (\texttt{SyntheticCube}) to a rough \ac{SED} (\texttt{SyntheticSED}). Additionally, the \textsc{FluxCompensator} provides an algorithm to extract a photometric flux within wavelength bins from the 3-d spectral cube (\texttt{SyntheticCube}). The resulting object resembles a \texttt{SyntheticFlux}, but obtained without filter convolution. Furthermore, during the pipeline one can extract also the flux within certain wavelength bins: the resulting object is a \texttt{SyntheticFlux} object. Furthermore, for the object \texttt{SyntheticImage} the total fluxes can be extracted any time. 

The \textsc{FluxCompensator} can produce a variety of outputs. With one line of Python code, users can output the resulting image to a \acs{FITS} file, a pre-formatted plot or one can plot \acp{SED} and photometric fluxes together. This can be done at any stage of the pipeline and also when synthetic interfaces are used.

It is also possible to create multi-color images or movies of different inclinations or time-steps as it was already possible for the \textsc{Hyperion} outputs. For further information visit \url{http://www.hyperion-rt.org} or \url{https://github.com/koepferl/FluxCompensator} and \url{http://FluxCompensator.readthedocs.io}.

\newpage
\subsection{Further Features}
Additional tools and features are provided by the \textsc{FluxCompensator} and are described below:

\subsubsection{Unit Conversion}
The \textsc{FluxCompensator} can convert fluxes, flux densities and surface brightnesses (see Section~\ref{C2:virt}) between the following units:\\
\begin{itemize}
\item \si[mode=text]{ergs.s^{-1}.cm^{-2}}\\[-0.5cm]
\item \si[mode=text]{ergs.s^{-1}.cm^{-2}.Hz^{-1}}\\[-0.5cm]
\item \si{\Jy}\\[-0.5cm]
\item \si{\milli\Jy}\\[-0.5cm]
\item \si{\mega\Jy\per\sr}\\[-0.5cm]
\item \si{Jy.arcsec^{-2}}
\end{itemize}
It is further possible for scalar fluxes (e.\,g.~as member of \texttt{SyntheticFlux}) to convert to magnitudes. For the filters in the database we list the required zero-magnitude flux. Otherwise this information should be specified by the user. For the zero-magnitude flux of the detectors included in the built-in database and their references, see Table~\ref{C2:Appendix:Tab} in Appendix~\ref{C2:Appendix}.

\subsubsection{Spectral Transmission Curve Visualization}
The \textsc{FluxCompensator} provides a plotting tool to visualize the spectral transmission functions of chosen filters from the database and/or input filter curves. It is possible to compare transmission curves with different axis scaling and units.

\subsubsection{Stellar Field}
\begin{figure*}[t]
\includegraphics[width=\textwidth]{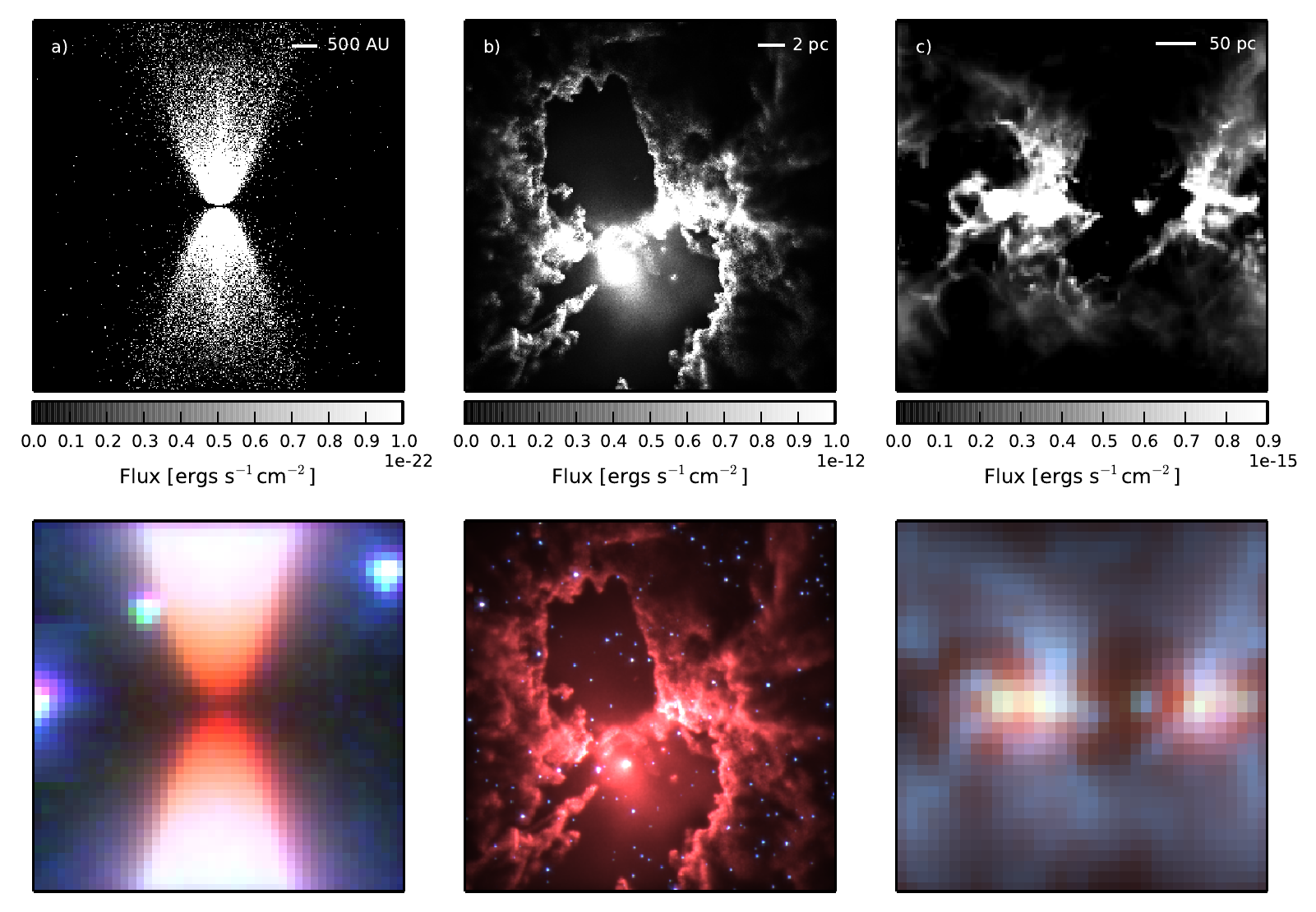}
\caption{In column a) a \ac{YSO} model at \SI{140}{\pc}, b) a star-forming region at \SI{10}{\kilo\pc} and c) the center of a galaxy at \SI{500}{\kilo\pc}. Top: Ideal single band synthetic observation at about \SI{8}{\microns} (a, b) and at about \SI{500}{\microns} (c) extracted directly from the radiative transfer output. Bottom: a,b) Realistic synthetic three-color images including field stars in \ac{GLIMPSE} colors and c) \emph{Herschel} three-color image. The maximum value of the stretch for the fluxes in \ac{IRAC} \SI{8}{\microns}, \ac{IRAC} \SI{4.5}{\microns}, \ac{IRAC} \SI{3.6}{\microns} (\acs{RGB}) are \SI[mode=text]{3.0e-14}{ergs.s^{-1}.cm^{-2}} (red), \SI[mode=text]{4.2e-14}{ergs.s^{-1}.cm^{-2}} (green) and \SI[mode=text]{1.3e-14}{ergs.s^{-1}.cm^{-2}} (blue) for Figure~\ref{C2:Fig:caseplot}\textcolor{lmugreen}{(a)} and \SI[mode=text]{1.2e-13}{ergs.s^{-1}.cm^{-2}} (red), \SI[mode=text]{2.0e-13}{ergs.s^{-1}.cm^{-2}} (green) and \SI[mode=text]{8.0e-14}{ergs.s^{-1}.cm^{-2}} (blue) for Figure~\ref{C2:Fig:caseplot}\textcolor{lmugreen}{(b)}. The maximum value of the stretch for the fluxes in \ac{SPIRE} \SI{500}{\microns}, \ac{PACS} \SI{160}{\microns}, \ac{PACS} \SI{70}{\microns} (\acs{RGB}) in for Figure~\ref{C2:Fig:caseplot}\textcolor{lmugreen}{(c)} are\SI[mode=text]{3.0e-14}{ergs.s^{-1}.cm^{-2}} (red), \SI[mode=text]{2.8e-13}{ergs.s^{-1}.cm^{-2}} (green) and \SI[mode=text]{9.3e-15}{ergs.s^{-1}.cm^{-2}} (blue).}
\label{C2:Fig:caseplot}
\end{figure*}

For a given 2-d image (\texttt{SyntheticImage}) the \textsc{FluxCompensator} can add a given number of foreground and background stars to the image and can deal with the extinction by the density distribution in the model (for the background stars) and \ac{PSF} convolution effects. For an illustration of this, see Figure~\ref{C2:Fig:caseplot}. To create a realistic stellar field, this feature of the \textsc{FluxCompensator} can be used in combination with stellar population synthesis models \citep[e.\,g.][]{Robin:2003}. 

\section{Compatibility with Other Codes}
\label{C2:comp}
The \textsc{FluxCompensator} package has been designed to be able to post-process the radiative transfer output of \textsc{Hyperion}. It follows the same philosophy as \textsc{Hyperion} in using object-oriented Python scripts and making it therefore more user friendly and generic. Very little Python knowledge is required to work with either \textsc{Hyperion} or the \textsc{FluxCompensator}. As the tool has been designed for \textsc{Hyperion}, some features of the \textsc{FluxCompensator} will only work optimally if \textsc{Hyperion} inputs are used. 
However, in principle it should be easy to adapt the \textsc{FluxCompensator} so that it can read in any arbitrary ideal observations from other radiative transfer codes (e.\,g.~those listed in Section~\ref{C2:intro}), provided that they can output a spectral cube. Aside from the input, most of the code is not specific to \textsc{Hyperion}. 

\section{Possible Applications}
\label{C2:case}

We show additional features of the \textsc{FluxCompensator} in this section and produce realistic synthetic observations of a \ac{YSO} model, a star-forming region and a center of a galaxy. This time we will not combine the realistic synthetic observation with a real observation as was done in Figure~\ref{C2:Fig:add2obs}. We rather produce three color images and additionally use the stellar field option to include reddened field stars to two of the synthetic observations to make them even more realistic.

In the case of the \ac{YSO}, we used the built-in \ac{YSO} toy model setup in \textsc{Hyperion} to construct the density distribution (see Figure~\ref{C2:Fig:density}\textcolor{lmugreen}{(a)}). For the star-forming region we used the density distribution displayed in Figure~\ref{C2:Fig:density}\textcolor{lmugreen}{(b)}, which was calculated from hydrodynamical simulations by \cite{DaleIoni:2012}. For the edge-on view of a center of a galaxy we used the dust temperature including gas-to-dust heating effects, which was calculated within the \textsc{SILCC} collaboration \citep{Girichidis:2015,Walch:2014}, to produce synthetic observations. The density distribution of the galaxy is shown in Figure~\ref{C2:Fig:density}\textcolor{lmugreen}{(c)}.

With the \textsc{FluxCompensator}, we extract three-color images in the colors of the \ac{GLIMPSE} survey displayed in Figure~\ref{C2:Fig:caseplot}\textcolor{lmugreen}{(a,b)}. The ideal synthetic observations (at about \SI{8}{\microns}) of the realistic synthetic observations are also plotted in Figure~\ref{C2:Fig:caseplot}\textcolor{lmugreen}{(a,b)}. For the central region of a galaxy we show a three-color \ac{Hi-GAL} image in Figure~\ref{C2:Fig:caseplot}\textcolor{lmugreen}{(c)}. In this case, field stars are not added. We plot the ideal synthetic observation (at about \SI{500}{\microns}) from \textsc{Hyperion}.\\

Realistic synthetic observations (as in Figure~\ref{C2:Fig:pipeline}, Figure~\ref{C2:Fig:add2obs} and Figure~\ref{C2:Fig:caseplot}) can help to critically test tools used by observers (e.\,g.~structure finding mechanisms, modified blackbody fitting) to extract physical quantities (e.\,g.~\acp{SFR}, gas masses) from real observations. \\

In \cite{Koepferl:2015} we found that the \ac{SFR} in the \ac{CMZ} of the Milky Way was overestimated by at least a factor of 3 by \cite{Yusef:2009}, who derived the \ac{SFR} directly from \acp{YSO} detected at \SI{24}{\microns}. Aided by the \textsc{FluxCompensator} in \cite{Koepferl:2015} we showed that the misclassified \acp{YSO}, which lead to the overestimation of the rate, could also be main-sequence stars or other more evolved objects in an ambient medium. We provided classification criteria which will help in future when classifying \acp{YSO} in this region.

In \cite{KDR1:inprep} we went one step further created realistic synthetic observations of \cite{DaleBoth:2014} simulations at different distances, angles and time-steps with the \textsc{FluxCompensator}. With this set of realistic synthetic observations we tested the accuracy of techniques which are commonly used to estimate the \ac{SFR}  (\citealt{KDR2b:inprep} and \citeauthor[][in prep.]{KDR3:inprep}) and the gas mass, dust surface densities and dust temperatures \citep{KDR2a:inprep} of star-forming regions.\\

Already existing realistic synthetic observations of star-forming regions in SPITZER and HERSCHEL bands have been already published in several data releases: \url{https://doi.org/10.5281/zenodo.260106}, \url{https://doi.org/10.5281/zenodo.56424}, \url{https://doi.org/10.5281/zenodo.31293}, \url{https://doi.org/10.5281/zenodo.31294}.

\section{Future Implementations \& Software}
\label{C2:futur}
With the \textsc{FluxCompensator}, as with \textsc{Hyperion}, we are still working on new tools to extend and improve the package in a continuous process. We are currently working on the following new applications of the \textsc{FluxCompensator}:
\begin{itemize}
\item Wavelength-dependent \ac{PSF} convolution\\[-0.5cm]
\item Noise standard deviation estimation directly from the information provided about the specific detector\\[-0.5cm] 
\item Saturation limits\\[-0.5cm]    
\item Gas radiative transfer post-processing\\[-0.5cm]
\item Synthetic interferometry \& line observations e.\,g.~by building interfaces to the \ac{CASA}\\[-0.5cm]
\item Astropy affiliated package
\end{itemize}

The \textsc{FluxCompensator} package has been designed to be able to post-process the radiative transfer output of \textsc{Hyperion} \citep{Robitaille:2011} and is extendable to other radiative transfer codes. Further, this research made use of Astropy, a community-developed core Python package for Astronomy \citep{Astropy:2013}, matplotlib, a Python plotting library \citep{Hunter:2007}, Pytest, test writing software for Python codes \citep{pytest}, Scipy, an open source scientific computing tool \citep{Scipy}, the NumPy package \citep{NumPy} and IPython, an interactive Python application \citep{IPython}.

\section{Summary}
\label{C2:summary} 
We have presented the \textsc{FluxCompensator} tool which can compute realistic synthetic observations from ideal radiative transfer observations, accounting for pixel resolution, reddening, \ac{PSF} and filter convolution and noise. Taking into account these effects is important when modeling distant astronomical objects such as \acp{YSO}, star-forming regions or galaxies, where for instance multiple objects can be blended into a single source. 

The \textsc{FluxCompensator} can be used to gauge and construct observational techniques, which extract properties of the observed objects or classify them. 
In \cite{Koepferl:2015}, \cite{KDR1:inprep}, we use the \textsc{FluxCompensator} to produce realistic synthetic observations to test estimates of \acp{SFR} and gas and dust gas properties in star-forming regions in \cite{KDR2a:inprep,KDR2b:inprep} and \citeauthor[][(in prep.)]{KDR3:inprep}.

The tool is now publicly available. For more information visit the \textsc{FluxCompensator} webpage \url{https://github.com/koepferl/FluxCompensator}, \url{https://doi.org/10.5281/zenodo.815629} and \url{http://FluxCompensator.readthedocs.io}.

\section{Acknowledgements}
We thank the referee for a constructive report that helped us improve the clarity and the strength of the results presented in our paper. This work was carried out in the Max Planck Research Group \textit{Star formation throughout the Milky Way Galaxy} at the Max Planck Institute for Astronomy. C.K. is a fellow of the International Max Planck Research School for Astronomy and Cosmic Physics (IMPRS) at the University of Heidelberg, Germany and acknowledges support. C.K. also acknowledges support from the and the Bayerischen Gleichstellungsf\"orderung. We would like to thank Jim Dale and Steffi Walch for the provided simulations. 

\bibliographystyle{apj}
\bibliography{paper_FC}

\clearpage
\appendix

\section{Database Overview}
\label{C2:Appendix}

Here we provide the information, which the \textsc{FluxCompensator} includes in the built-in database. The corresponding references are given in the footnotes.
\input{tab01}

\input{header}

\end{document}

%% file: tab01.tex
\begin{table}[b]
\caption[Information about Telescopes and Detectors]{\label{C2:Appendix:Tab}Information about Telescopes and Detectors Provided in the \textsc{FluxCompensator} Database}
			\begin{tabular*}{\textwidth}{p{1.3cm}r|p{1cm}D{.}{.}{4}|lD{.}{.}{3}D{.}{.}{3}D{.}{.}{3}rr|cD{.}{.}{3}c}
\hline\\[-13pt]
\hline\\[-5pt]
&&&&&&&&&&&&\\
				\multicolumn{2}{c|}{telescope}	   &   \multicolumn{2}{c|}{detector}	       &       \multicolumn{6}{c|}{filter}	       &  \multicolumn{3}{c}{PSF\footnotemark[1]}     \\
				\multicolumn{1}{c}{\multirow{2}{*}{name}}		&
				\multicolumn{1}{c|}{diameter}					&
				\multicolumn{1}{c}{\multirow{2}{*}{name}}		&
				\multicolumn{1}{c|}{zero-point}					& 
				\multicolumn{1}{c}{\multirow{2}{*}{origin}}		&
				\multicolumn{3}{c}{wavelength [$\mu$m]}  		&
				\multicolumn{1}{c}{\multirow{2}{*}{$\alpha$}}	&
				\multicolumn{1}{c|}{\multirow{2}{*}{$\beta$}}	&
				\multicolumn{1}{c}{\multirow{2}{*}{built-in}}		&
				\multicolumn{1}{c}{\multirow{1}{*}{resolution}}  &
				\multicolumn{1}{c}{\multirow{2}{*}{sampled}}\\	
				
				&
				\multicolumn{1}{c|}{[cm]}			&
				&
				\multicolumn{1}{c|}{[Jy]}	&
				&
				\multicolumn{1}{c}{center}			&
				\multicolumn{1}{c}{min}        		&
				\multicolumn{1}{c}{max}				&
				&
				&
				&
				\multicolumn{1}{c}{[$\frac{\mbox{arcsec}}{\mbox{pixel}}$]}	&
\\[6pt]
 				\hline\\[-5pt]
2MASS
&130.\footnotemark[2]
&\verb|J[1.2]|
&1594.\footnotemark[3]\footnotemark[4]
&\footnotemark[3]\footnotemark[4]
&1.235\footnotemark[3]\footnotemark[4]
&1.062\footnotemark[3]
&1.450\footnotemark[3]
&1
&0
&\multicolumn{1}{c}{-}
&\multicolumn{1}{c}{-}
&\multicolumn{1}{c}{-}\\

2MASS
&130.\footnotemark[2]
&\verb|H[1.7]|
&1024.\footnotemark[3]\footnotemark[4]
&\footnotemark[3]\footnotemark[4]
&1.662\footnotemark[3]\footnotemark[4]
&1.289\footnotemark[3]
&1.914\footnotemark[3]
&1
&0
&\multicolumn{1}{c}{-}
&\multicolumn{1}{c}{-}
&\multicolumn{1}{c}{-}\\

2MASS
&130.\footnotemark[2]
&\verb|K[2.2]|
&666.7\footnotemark[3]\footnotemark[4]
&\footnotemark[3]\footnotemark[4]
&2.159\footnotemark[3]\footnotemark[4]
&1.900\footnotemark[3]
&2.399\footnotemark[3]
&1
&0
&\multicolumn{1}{c}{-}
&\multicolumn{1}{c}{-}
&\multicolumn{1}{c}{-}\\[6pt]

SPITZER
&85.\footnotemark[5]
&\verb|IRAC[3.5]|
&280.9\footnotemark[6]\footnotemark[7]
&\footnotemark[8]\footnotemark[9]
&3.550\footnotemark[10]
&3.081\footnotemark[8]
&4.010\footnotemark[8]
&1\footnotemark[10]\footnotemark[7]\footnotemark[11]
&0\footnotemark[10]\footnotemark[7]\footnotemark[11]
&x
&1.221
&4\\

SPITZER
&85.\footnotemark[5]
&\verb|IRAC[4.5]|
&179.7\footnotemark[6]\footnotemark[7]
&\footnotemark[8]\footnotemark[9]
&4.493\footnotemark[10]
&3.722\footnotemark[8]
&5.222\footnotemark[8]
&1\footnotemark[10]\footnotemark[7]\footnotemark[11]
&0\footnotemark[10]\footnotemark[7]\footnotemark[11]
&x
&1.213
&4\\

SPITZER
&85.\footnotemark[5]
&\verb|IRAC[5.7]|
&115.0\footnotemark[6]\footnotemark[7]
&\footnotemark[8]\footnotemark[9]
&5.731\footnotemark[10]
&4.744\footnotemark[8]
&6.623\footnotemark[8]
&1\footnotemark[10]\footnotemark[7]\footnotemark[11]
&0\footnotemark[10]\footnotemark[7]\footnotemark[11]
&x
&1.222
&4\\

SPITZER
&85.\footnotemark[5]
&\verb|IRAC[8]|
&64.9\footnotemark[6]
&\footnotemark[8]\footnotemark[9]
&7.872\footnotemark[10]
&6.151\footnotemark[8]
&10.497\footnotemark[8]
&1\footnotemark[10]\footnotemark[7]\footnotemark[11]
&0\footnotemark[10]\footnotemark[7]\footnotemark[11]
&x
&1.220
&4\\[6pt]

SPITZER
&85.\footnotemark[5]
&\verb|MIPS[24]|
&7.17\footnotemark[12]
&\footnotemark[13]
&23.68\footnotemark[12]
&18.005\footnotemark[13]
&32.307\footnotemark[13]
&-2\footnotemark[14]
&-1
&x
&2.49
&5\\

SPITZER
&85.\footnotemark[5]
&\verb|MIPS[70]|
&0.778\footnotemark[12]
&\footnotemark[13]
&71.42\footnotemark[12]
&49.960\footnotemark[13]
&111.022\footnotemark[13]
&-2\footnotemark[14]
&-1
&x
&9.85
&5\\

SPITZER
&85.\footnotemark[5]
&\verb|MIPS[160]|
&0.159\footnotemark[12]
&\footnotemark[13]
&155.9\footnotemark[12]
&100.085\footnotemark[13]
&199.92\footnotemark[13]
&-2\footnotemark[14]
&-1
&x
&16.
&5\\[6pt]

IRAS
&57.\footnotemark[15]
&\verb|IRAS[12]|
&30.88\footnotemark[16]
&\footnotemark[17]
&12.\footnotemark[17]
&7.0\footnotemark[17]
&15.5\footnotemark[17]
&1\footnotemark[18]
&-1
&\multicolumn{1}{c}{-}
&\multicolumn{1}{c}{-}
&\multicolumn{1}{c}{-}\\

IRAS
&57.\footnotemark[15]
&\verb|IRAS[27]|
&7.26\footnotemark[16]
&\footnotemark[17]
&25.\footnotemark[17]
&16.0\footnotemark[17]
&31.5\footnotemark[17]
&1\footnotemark[18]
&-1
&\multicolumn{1}{c}{-}
&\multicolumn{1}{c}{-}
&\multicolumn{1}{c}{-}\\

IRAS
&57.\footnotemark[15]
&\verb|IRAS[60]|
&1.11\footnotemark[16]
&\footnotemark[17]
&60.\footnotemark[17]
&27.0\footnotemark[17]
&87.0\footnotemark[17]
&1\footnotemark[18]
&-1
&\multicolumn{1}{c}{-}
&\multicolumn{1}{c}{-}
&\multicolumn{1}{c}{-}\\

IRAS
&57.\footnotemark[15]
&\verb|IRAS[100]|
&0.39\footnotemark[16]
&\footnotemark[17]
&100.\footnotemark[17]
&65.0\footnotemark[17]
&140.0\footnotemark[17]
&1\footnotemark[18]
&-1
&\multicolumn{1}{c}{-}
&\multicolumn{1}{c}{-}
&\multicolumn{1}{c}{-}\\[6pt]

HERSCHEL
&350.\footnotemark[19]
&\verb|PACS[70]|
&0.78\footnotemark[16]
&\footnotemark[20]
&70.\footnotemark[21]
&48.721\footnotemark[20]
&157.480\footnotemark[20]
&1\footnotemark[22]
&-1
&x
&4.
&10\\

HERSCHEL
&350.\footnotemark[19]
&\verb|PACS[100]|
&0.38\footnotemark[16]
&\footnotemark[20]
&100.\footnotemark[21]
&48.960\footnotemark[20]
&186.916\footnotemark[20]
&1\footnotemark[22]
&-1
&x
&4.
&10\\

HERSCHEL
&350.\footnotemark[19]
&\verb|PACS[160]|
&0.14\footnotemark[16]
&\footnotemark[20]
&160.\footnotemark[21]
&105.263\footnotemark[20]
&500.000\footnotemark[20]
&1\footnotemark[22]
&-1
&x
&4.
&10\\[6pt]

HERSCHEL
&350.\footnotemark[19]
&\verb|SPIRE[250]|
&0.06\footnotemark[16]
&\footnotemark[23]
&250.\footnotemark[19]
&115.015\footnotemark[23]
&291.411\footnotemark[23]
&1\footnotemark[24]
&-1
&x
&6.
&10\\

HERSCHEL
&350.\footnotemark[19]
&\verb|SPIRE[350]|
&0.03\footnotemark[16]
&\footnotemark[23]
&350.\footnotemark[19]
&137.397\footnotemark[23]
&419.451\footnotemark[23]
&1\footnotemark[24]
&-1
&x
&6.
&10\\

HERSCHEL
&350.\footnotemark[19]
&\verb|SPIRE[500]|
&0.01\footnotemark[16]
&\footnotemark[23]
&500.\footnotemark[19]
&316.429\footnotemark[23]
&603.042\footnotemark[23]
&1\footnotemark[24]
&-1
&x
&6.
&10\\[6pt]

WISE
&40.\footnotemark[25]
&\verb|WISE[3.4]|
&309.540\footnotemark[26]
&\footnotemark[27]\footnotemark[28]
&3.353\footnotemark[26]
&2.53\footnotemark[28]
&6.50\footnotemark[28]
&2\footnotemark[29]
&0
&\multicolumn{1}{c}{-}
&\multicolumn{1}{c}{-}
&\multicolumn{1}{c}{-}\\

WISE
&40.\footnotemark[25]
&\verb|WISE[4.6]|
&171.787\footnotemark[26]
&\footnotemark[27]\footnotemark[28]
&4.603\footnotemark[26]
&2.53\footnotemark[28]
&8.00\footnotemark[28]
&2\footnotemark[29]
&0
&\multicolumn{1}{c}{-}
&\multicolumn{1}{c}{-}
&\multicolumn{1}{c}{-}\\

WISE
&40.\footnotemark[25]
&\verb|WISE[12]|
&31.674\footnotemark[26]
&\footnotemark[27]\footnotemark[28]
&11.562\footnotemark[26]
&2.53\footnotemark[28]
&28.55\footnotemark[28]
&2\footnotemark[29]
&0
&\multicolumn{1}{c}{-}
&\multicolumn{1}{c}{-}
&\multicolumn{1}{c}{-}\\

WISE
&40.\footnotemark[25]
&\verb|WISE[22]|
&8.363\footnotemark[26]
&\footnotemark[27]\footnotemark[28]
&22.088\footnotemark[26]
&2.53\footnotemark[28]
&28.55\footnotemark[28]
&2\footnotemark[29]
&0
&\multicolumn{1}{c}{-}
&\multicolumn{1}{c}{-}
&\multicolumn{1}{c}{-}\\[6pt]
 				\hline\\[-5pt]

\end{tabular*}
\footnotetext[1]{PSFs origin: \url{http://dirty.as.arizona.edu/~kgordon/mips/conv_psfs/conv_psfs.html}}

\footnotetext[2]{2MASS Handbook: \url{http://www.ipac.caltech.edu/2mass/releases/first/doc/sec3_1a.html}}
\footnotetext[3]{2MASS Handbook: \url{http://www.ipac.caltech.edu/2mass/releases/allsky/doc/sec6_4a.html}}
\footnotetext[4]{\cite{Cohen:2003}}

\footnotetext[5]{Spitzer Handbook: \\\url{http://irsa.ipac.caltech.edu/data/SPITZER/docs/spitzermission/missionoverview/spitzertelescopehandbook/13/}}
\footnotetext[6]{IRAC Handbook: \url{http://irsa.ipac.caltech.edu/data/SPITZER/docs/irac/iracinstrumenthandbook/17/}}
\footnotetext[7]{\cite{Reach:2005}}
\footnotetext[8]{IRAC Handbook: \url{http://irsa.ipac.caltech.edu/data/SPITZER/docs/irac/calibrationfiles/spectralresponse/}}
\footnotetext[9]{\cite{Quijada:2004}}
\footnotetext[10]{IRAC Handbook: \url{http://irsa.ipac.caltech.edu/data/SPITZER/docs/irac/iracinstrumenthandbook/18/}}
\footnotetext[11]{\cite{Hora:2008}}

\footnotetext[12]{MIPS Handbook: \url{http://irsa.ipac.caltech.edu/data/SPITZER/docs/mips/mipsinstrumenthandbook/49/}}
\footnotetext[13]{MIPS Handbook: \url{http://irsa.ipac.caltech.edu/data/SPITZER/docs/mips/calibrationfiles/spectralresponse/}}
\footnotetext[14]{MIPS Handbook: \url{http://irsa.ipac.caltech.edu/data/SPITZER/docs/mips/mipsinstrumenthandbook/51/}}

\footnotetext[15]{IRAS Handbook: \url{http://irsa.ipac.caltech.edu/IRASdocs/iras_mission.html}}
\footnotetext[16]{Filter information: \url{http://svo2.cab.inta-csic.es/theory/fps/index.php}}
\footnotetext[17]{IRAS Handbook: \url{http://irsa.ipac.caltech.edu/IRASdocs/exp.sup/ch2/tabC5.html}}
\footnotetext[18]{IRAS Handbook: \url{http://lambda.gsfc.nasa.gov/product/iras/docs/exp.sup/ch6/C3.html}}

\footnotetext[19]{Herschel Handbook: \url{http://herschel.esac.esa.int/Docs/SPIRE/html/}}
\footnotetext[20]{PACS Handbook: \url{https://nhscsci.ipac.caltech.edu/sc/index.php/Pacs/FilterCurves}}
\footnotetext[21]{PACS Handbook: \url{http://herschel.esac.esa.int/Docs/PACS/html/ch03s02.html}}
\footnotetext[22]{PACS Handbook: \url{http://herschel.esac.esa.int/Docs/PACS/html/ch03s03.html}}

\footnotetext[23]{provided by the \emph{Herschel} helpdesk}
\footnotetext[24]{SPIRE Handbook: \url{http://herschel.esac.esa.int/hcss-doc-11.0/load/spire_drg/html/ch05s07.html}}

\footnotetext[25]{WISE Handbook: \url{http://wise2.ipac.caltech.edu/docs/release/allsky/expsup/sec3_2.html}}
\footnotetext[26]{\cite{Jarrett:2011}}
\footnotetext[27]{\cite{Wright:2010}}
\footnotetext[28]{WISE Handbook: \url{http://wise2.ipac.caltech.edu/docs/release/prelim/expsup/sec4_3g.html}}
\footnotetext[29]{WISE Handbook: \url{http://wise2.ipac.caltech.edu/docs/release/allsky/expsup/sec4_4h.html}}
\end{table}

%% file: header.tex
\begin{acronym}[ATLASGAL]
\acro{2MASS}{Two Micron All-Sky Survey}
\acro{AGB}{Asymptotic Giant Branch}
\acro{ALMA}{Atacama Large Millimeter/Submillimeter Array}
\acro{AMR}{Adaptive Mesh Refinement}
\acro{ATLASGAL}{APEX Telescope Large Area Survey of the Galaxy}
\acro{BGPS}{Bolocam Galactic Plane Survey}
\acro{c1}{sample clipping of only neutral particles within a box of \SI{30}{\pc}}
\acro{c2}{sample clipping of \acs{c1} particles and for a certain threshold temperature}
\acro{c2d}{Cores to Disks Legacy}
\acro{CASA}{Common Astronomy Software Applications package}
\acro{cm}{centimeter}
\acro{CMF}{core mass function}
\acro{CMZ}{central molecular zone}
\acro{D1}{distance at \SI{3}{\kpc}}
\acro{D14}{\acs{SPH} simulations performed by Jim Dale and collaborators \citep{DaleI:2011,DaleIoni:2012,DaleIoni:2013,DaleWind:2013,DaleBoth:2014}}
\acro{D2}{distance at \SI{10}{\kpc}}
\acro{DT1}{temperature coupling of radiative transfer \& hydrodynamical temperature}
\acro{DT2}{temperature coupling of the radiative transfer \& isothermal temperature}
\acro{DT3}{no temperature coupling of the radiative transfer temperature}
\acro{e1}{using the \cite{Ulrich:1976} envelope profile to extrapolate the envelope inwards}
\acro{e2}{using the \cite{Ulrich:1976} envelope profile with suppressed singularity to extrapolate the envelope inwards}
\acro{e3}{using a power-law envelope profile to extrapolate the envelope inwards}
\acro{EOS}{equation of state}
\acro{FIR}{far-infrared}
\acro{FITS}{Flexible Image Transport System}
\acro{FWHM}{full-width at half-maximum}
\acro{GLIMPSE}{Galactic Legacy Infrared Mid-Plane Survey Extraordinaire}
\acro{GMC}{Giant Molecular Clouds}
\acro{Hi-GAL}{\emph{Herschel} Infrared Galactic Plane Survey}
\acro{HST}{\emph{Hubble} Space Telescope}
\acro{HWHM}{half-width at half-maximum}
\acro{IMF}{initial mass function}
\acro{IR}{infrared}
\acro{IRAC}{Infrared Array Camera}
\acro{IRAS}{Infrared Astronomical Satellite}
\acro{ISM}{interstellar medium}
\acro{$K$}{K band}
\acro{LTE}{local thermodynamical equilibrium}
\acro{MIPS}{Multiband Imaging Photometer for \emph{Spitzer}}
\acro{MIPSGAL}{\acs{MIPS} Galactic Plane Survey}
\acro{MIR}{mid-infrared}
\acro{MS}{main-sequence}
\acro{mm}{millimeter}
\acro{NASA}{National Aeronautics and Space Administration}
\acro{NIR}{near-infrared}
\acro{N-PDF}{column density \acs{PDF}}
\acro{O1}{xy plane}
\acro{O2}{xz plane}
\acro{O3}{yz plane}
\acro{p1}{parameter evaluation version from \acs{SPH} kernel function}
\acro{p2}{parameter evaluation version from \acs{SPH} splitted kernel distribution}
\acro{p3}{parameter evaluation version from \acs{SPH} random distribution}
\acro{PACS}{Photoconductor Array Camera and Spectrometer}
\acro{PAH}{polycyclic aromatic hydrocarbon}
\acro{PDF}{probability distribution function}
\acro{PDR}{Photon Dominated Region}
\acro{PSF}{point-spread-function}
\acro{px}{one of the parameter evaluation version \acs{p1}, \acs{p2}, \acs{p3}}
\acro{RGB}{red, green and blue}
\acro{CM1}{circumstellar setup with background density and sink mass as stellar mass}
\acro{CM2}{circumstellar setup by a toy model with \acs{e2} envelope superposition density and corrected stellar mass, protoplanetary disk and envelope cavity}
\acro{CM3}{circumstellar setup by a toy model with \acs{e3} envelope superposition density and corrected stellar mass, protoplanetary disk and envelope cavity}
\acro{s1}{Voronoi site placement version at \acs{SPH} particle position}
\acro{s2}{Voronoi site placement version as \acs{s1} including sites at sink particles}
\acro{s3}{Voronoi site placement version as \acs{s2} including circumstellar sites}
\acro{SAO}{Smithsonian Astrophysical Observatory}
\acro{SED}{spectral energy distribution}
\acro{SFE}{star-formation efficiency}
\acro{SFR}{star-formation rate}
\acro{SFR24}{technique to measure the \acs{SFR} using the \SI{24}{\microns} tracer}
\acro{SFR70}{technique to measure the \acs{SFR} using the \SI{70}{\microns} tracer}
\acro{SFRIR}{technique to measure the \acs{SFR} using the total infrared tracer}
\acro{Sgr}{Sagittarius}
\acro{SMA}{Submillimeter Array}
\acro{SPH}{smooth particle hydrodynamics}
\acro{SPIRE}{Spectral and Photometric Imaging Receiver}
\acro{sub-mm}{sub-millimeter}
\acro{UKIDSS}{UKIRT Infrared Deep-Sky Survey}
\acro{UKIRT}{UK Infrared Telescope}
\acro{UV}{ultra-violet}
\acro{WFCAM}{\acs{UKIRT} Wide Field Camera}
\acro{WISE}{Wide-field Infrared Survey Explorer}
\acro{YSO}{young stellar object}
\end{acronym}